\documentstyle[11pt,flushrt,palatino,amssym,epsf]{nature}

\def\apj{Astrophys. J.}
\def\apjl{Astrophys. J.}
\def\aj{Astron. J.}
\def\mnras{Mon. Not. R. Astron. Soc.}

\def\nature{Nature}
\def\science{Science}

\def\spose#1{\hbox to 0pt{#1\hss}}
\def\lta{\mathrel{\spose{\lower 3pt\hbox{$\mathchar"218$}}
     \raise 2.0pt\hbox{$\mathchar"13C$}}}
\def\gta{\mathrel{\spose{\lower 3pt\hbox{$\mathchar"218$}}
     \raise 2.0pt\hbox{$\mathchar"13E$}}}

\def\sun{\odot}
\def\check{\checkmark}
\def\omegam{{\Omega_{\rm m}}}
\def\omegab{{\Omega_{\rm b}}}
\def\omegal{{\Omega_\Lambda}}

\begin{document}

\title{\sc Do globular clusters time the Universe?}

\author{Oleg Y. Gnedin, Ofer Lahav, Martin J. Rees \\
        Institute of Astronomy, Cambridge CB3 0HA}

\maketitle

{\bf Estimating the age of the Universe is an old problem.  Rapid
progress in observational cosmology in recent years has led to more
accurate values of the fundamental parameters.  The current most popular
model is a flat Universe, with about 30\% of the critical density in the
form of matter (baryonic and non-baryonic) and 70\% in the form of `dark
energy'.  These two densities, together with the Hubble constant
(estimated to be about 70 km s$^{-1}$ Mpc$^{-1}$), constrain the age of
the Universe to be approximately 13 Gyr.  This expansion age is
uncomfortably close to the age of the oldest globular clusters
(approximately 12.5 Gyr), in particular if they formed relatively
recently.  We review here proposed models of globular cluster formation
and point out possible conflicts with cosmology.}

In a Big Bang model with a cosmological constant $\Lambda$ the age of
the Universe is determined by three present-epoch parameters: the Hubble
constant $H_0 = 100 \, h$ km s$^{-1}$ Mpc$^{-1}$ = (9.78 Gyr$)^{-1} \,
h$, the mass density parameter $\omegam$, and the scaled cosmological
constant $\omegal \equiv \Lambda/(3 H_0^2)$.  The cosmological constant
tends to stretch the age of the Universe, but if the dark energy were in
the form of 'quintessence' which decays with time, then $t_H$ would be
reduced.  For this reason (and to abbreviate our discussion) we focus on
the 'optimistic' case of constant $\Lambda$.  For a flat universe
($\omegam+\omegal =1$), supported by the recent Cosmic Microwave
Background (CMB) experiments, the age depends only on two free
parameters and is well approximated$^{\cite{P:99}}$ by:
\begin{equation}
 t_{H} \approx  \; {2 \over 3} H_0^{-1} \omegam^{-0.3}\;. 
\end{equation}
This gives us an insight to the way the errors propagate 
in the determination of the cosmic age:
\begin{equation}
  {\Delta t_{H} \over t_{H} } \approx 
  {\Delta H_0 \over H_0 } \; + \; 
  0.3{\Delta \omegam \over \omegam }\;.
\end{equation}
This shows that the fractional error in $H_0$ is three times as
important as the fractional error in $\omegam$.  Typically the quoted
error in $H_0$ is 10\%$^{\cite{Fetal:01}}$, while the recent range of
$\omegam$ is of order 50\%, so the expected fractional error in age is
25\% (e.g. for $t_{H} \approx 13$ Gyr, $\Delta t_{H} \approx 3$ Gyr).

Numerous studies$^{\cite{BOPS:99}\cite{Betal:01}\cite{WTZ:01}}$ have
compared and combined in a self-consistent way the most powerful cosmic
probes: the CMB, galaxy redshift surveys, galaxy cluster number counts,
type Ia supernovae, and galaxy peculiar velocities.  These studies
indicate that we live in a flat accelerating Universe, dominated by cold
dark matter (CDM) and `dark energy' (the cosmological constant $\Lambda$
or some generalization such as 'quintessence'$^{\cite{WCOS:00}}$).  More
precisely, the data are consistent with a $\Lambda$-CDM model with
$\omegam = 1 - \omegal \approx 0.3$ and $h \approx 0.7$, which
corresponds to an expansion age $t_{H} = 13.5$ Gyr.

While the above $\Lambda$-CDM model is currently very popular, there is
no simple theoretical explanation for the fortuitous near-equality of
the present-epoch matter density $\omegam$ and 'dark energy' density
$\omegal$, nor for the true nature of these components.  As a further
diagnostic, we would like to revive an old conundrum: is the age of the
Universe compatible with the ages of the oldest objects within it?  For
consistency, the ages of globular clusters ($t_{GC}$) must satisfy the
following relation with the epoch of their formation ($t_{f}$) and the
age of the Universe:
\begin{equation}
t_{H} = t_{f} + t_{GC}.
  \label{eq:tf}
\end{equation}

The age of the oldest GC is estimated to be $t_{GC} = 12.5 \pm 1.2$
Gyr$^{\cite{Ch:98}}$.  Radioactive dating of a very metal-poor star in
the Galaxy (using $^{238}$U) gives a similar age of $12.5 \pm 3$
Gyr$^{\cite{Cetal:01}}$.  Also, recently discovered Ly$\alpha$
emitters$^{\cite{Lya}}$ and Ly break galaxies$^{\cite{LBG}}$ at redshift
$z \sim 2-3$ show an already evolved stellar population as old as 1 Gyr,
comparable to the expansion age at that epoch.

We see that the ages of old objects might be uncomfortably close to the
age of expansion.  It is commonly assumed that $t_f \lta 2$
Gyr$^{\cite{Ch:98}}$, and hence $t_f$ is neglected in eq. (\ref{eq:tf}).
However, we point out here that $t_{f}$ is model dependent, and
different models predict a wide range of ages.

Formation of globular clusters in the Milky Way intimately relates to
the formation scenario of the Galaxy itself.  For decades, two
apparently conflicting models dominated the thought on Galaxy formation:
'monolithic collapse'$^{\cite{ELS:62}}$ versus highly fragmented star
formation$^{\cite{SZ:78}}$.  The realization that galaxies are embedded
into dark matter halos$^{\cite{OPY:74}}$ has led to a constructive
synthesis with cosmology.  According to the CDM model, galaxies form as
a result of gravitational growth and interaction of primordial
fluctuations.  Small objects collapse first and merge into larger
systems, extending the hierarchy to progressively higher masses.  The
present halo of the Milky Way formed in dozens of mergers of smaller
progenitors.

Our understanding of star formation has also improved dramatically.
Theoretical models still have many shortcomings$^{\cite{CBH:00}}$, but
the observational picture is becoming clearer.  It looks that most stars
form in clusters and associations of various sizes.  The hierarchy of
cluster masses ranges from the young OB associations to the old massive
globular clusters, with no special scale between 10 and $10^6\,
M_{\sun}$: $dN/dM \propto M^{-\alpha}$, $\alpha = 1.5 - 2
^{\cite{EE:97}\cite{WBM:00}}$.  On a global scale, the efficiency of
globular cluster formation remains low.  In large and small elliptical
galaxies, McLaughlin$^{\cite{M:99}}$ finds the same ratio of the mass of
globular clusters to the total baryonic (stellar + gaseous) mass of
their host galaxy, $\epsilon_{\sc gc} \equiv M_{\sc gc}/M_{\rm bar}
\approx 0.0026 \pm 0.0005$.

The observed distribution within galaxies and cluster mass function
$dN/dM$ differ from the {\it initial} ones due to dynamical evolution.
Small-mass clusters ($M < 10^5\, M_{\sun}$) are gradually destroyed by
stellar two-body relaxation and tidal interaction with the
Galaxy$^{\cite{S:87}\cite{GO:97}\cite{MW:97}\cite{GLO:99}}$, while very
massive clusters sink to the center via dynamical friction.  However,
most of the high-mass clusters are essentially unaffected by the
evolution, and therefore, preserve the shape of the initial mass
function.

\begin{figure}
\begin{center}
\epsfysize=12cm  \epsfbox{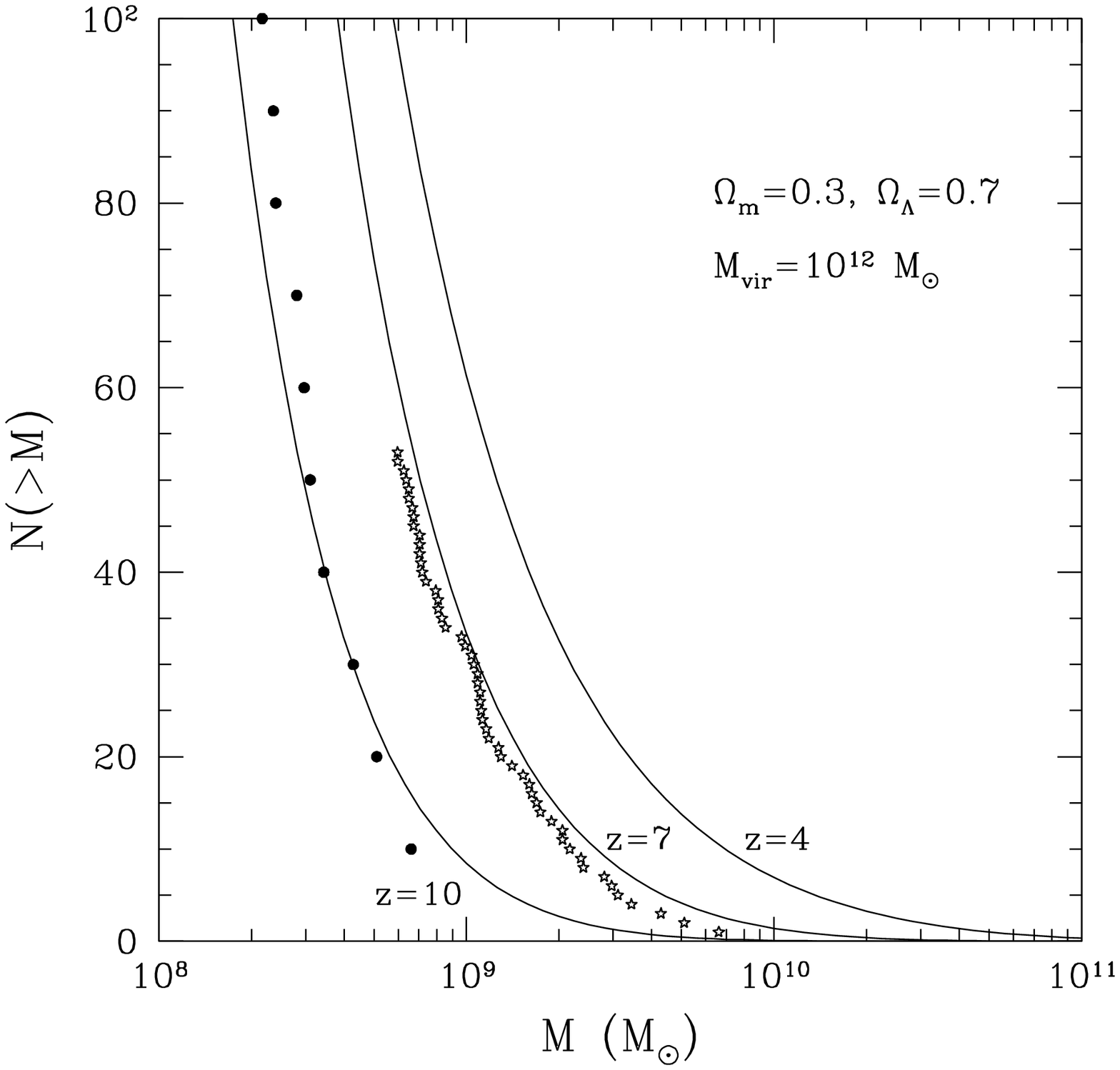}
\end{center}
\caption{In a hierarchical scenario that we propose, the oldest globular
clusters form in small progenitor galaxies with the mass of the
first-ranked cluster proportional to the amount of gas supply in the
progenitor, $f_b M$.  Massive clusters are almost unaffected by
dynamical evolution and serve as indicators of the initial distribution.
The plot compares the mass function of large globular clusters
(asterisks; data from \protect\cite{H:96}), renormalized by a factor
$(f_b \, \epsilon_{\sc gc})^{-1} \approx 3000$, with the cumulative
number of virialized progenitors of the present Milky Way halo.  For
three different redshifts, $N(>M)$ is calculated using the extended
Press-Schechter formalism$^{\protect\cite{LC:93}}$.  The two
distributions match at $z \approx 7$, so that the first clusters might
have formed soon thereafter.  The virialization epoch of the progenitors
is expected to have a statistical spread, which produces a corresponding
spread of globular cluster ages, in agreement with observations.  If the
efficiency of cluster formation $\epsilon_{\sc gc}$ is not universal,
i.e. significantly higher in some progenitors and lower in others, the
distribution may shift towards higher redshift.  For instance, filled
circles demonstrate the case where 10\% of the progenitors form clusters
with the efficiency $10 \, \epsilon_{\sc gc}$.  The expected epoch of
formation of oldest clusters thus follows to be $z_f \approx 7 - 10$.}
\label{fig:ggcs}
\end{figure}

The suggested scenarios of globular cluster formation can be grouped
into four main types, in decreasing order of the redshift of formation
$z_f$.  For reference, in our assumed cosmology ($\omegam=0.3,
\omegal=0.7, h=0.7$), redshifts $z_f = 7, 3, 1$ correspond to the
formation epochs $t_f = 0.8, 2.1$ and 5.8 Gyr, respectively.
\begin{itemize}
\item {\it Cosmological objects}

Peebles \& Dicke$^{\cite{PD:68}}$ were the first to propose that
globular clusters are cosmological objects formed soon after
recombination, with masses related to the Jeans mass (smallest mass of
gas clouds able to collapse).  However, including dark matter in the
calculation would reduce the mass of proto-globular clouds.  Also, the
expected efficiency of the conversion of gas into star clusters would be
much higher than the observed $\epsilon_{\sc gc}$.

\item {\it Hierarchical population} ($z_f \sim 7-10$)

A hierarchical formation is more promising within small galaxies --
progenitors of the Milky Way.  In progenitors just massive enough to be
cooled by atomic hydrogen, the gas collapses into a small disk where
gravitational instability causes fragmentation into clouds and formation
of star clusters of progressively larger sizes.  Each progenitor galaxy
may form a few globular clusters, still remaining very gas-rich before
merging into the Galactic halo.  Similar ideas have been put forward in
the literature$^{\cite{L:93}\cite{HP:94}\cite{MP:96}\cite{Cetal:00}}$.
See Figure \ref{fig:ggcs} for more detail.

\item {\it Large galactic halo} ($z_f \sim 1-3$)

Globular cluster formation as a result of thermal instability in hot gas
at virial temperature $T_{\rm vir} \sim 10^6$ K has been suggested by
Fall \& Rees$^{\cite{FR:85}}$ (variations of this idea include
$^{\cite{ML:92}\cite{VP:95}}$).  Dense clouds of metal-poor gas would
cool to $T_c \approx 10^4$ K, confined by the pressure of diffuse gas,
and would have the right Jeans mass to form massive globular clusters.
The epoch of formation is constrained to be $z_f \lta 3$, when halos
with large enough $T_{\rm vir}$ have been assembled.  On the other hand,
massive gaseous clouds must not have disrupted the thin disk of the
Galaxy (7-10 Gyr old), which demands $z_f \gta 1$.

\item {\it Mergers of disk galaxies} ($z_f \lta 1$)

Finally, the observations of young massive star clusters in colliding
galaxies prompted a ``merger'' scenario$^{\cite{AZ:92}}$.  In this
model, a merger of two spiral galaxies leads to a burst of star
formation producing a large population of metal-rich clusters, in
addition to the older population associated with the original spirals.
We include this model for completeness, although it has been originally
designed to account for the bimodal metallicity distribution of globular
clusters in elliptical galaxies and predicts that most of the clusters
form fairly recently.

\end{itemize}

In the table we provide a comparison of the formation models, as they
score against the observational properties of the Galactic globular
clusters.  The mark \check shows the model meets the observational
constraint, ? indicates possible but not certain agreement, and X
denotes clear disagreement with the data.  Hierarchical model seems to
fare better in this comparison, but scenarios with later formation are
still possible.

\begin{figure}
\begin{center}
\epsfysize=13cm  \epsfbox{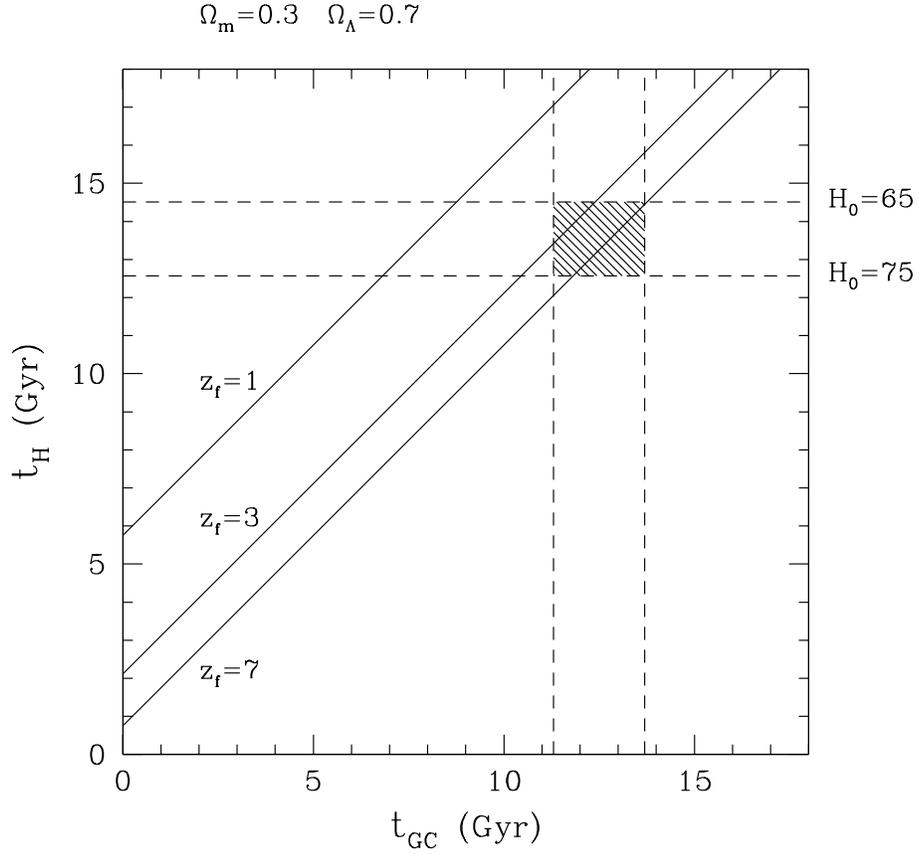}
\end{center}
\caption{The age of expansion vs the age of globular clusters, as a
function of the formation redshift $z_f$ (eq. \protect\ref{eq:tf}).
Shaded box contains the current values of $t_H$ and $t_{GC}$ within the
observational errors.  In a flat Universe with $\Omega_\Lambda = 1 -
\Omega_m = 0.7$, the age of expansion is $t_H = {2 \over 3} H_0^{-1} \;
\omegal^{-1/2} \ln [(1 + \omegal^{1/2}) (1-\omegal)^{-1/2}]$.}
\label{fig:age1}
\end{figure}

Figure \ref{fig:age1} illustrates eq. (\ref{eq:tf}) for the three
possible values of the formation redshift.  Early formation ($z_f = 7$)
is favorite, $z_f = 3$ is marginally consistent with the errors of $H_0$
and globular cluster ages, but any lower redshift is ruled out in the
flat Universe model with $\Omega_m = 0.3$.

\begin{figure}
\begin{center}
\epsfysize=13cm  \epsfbox{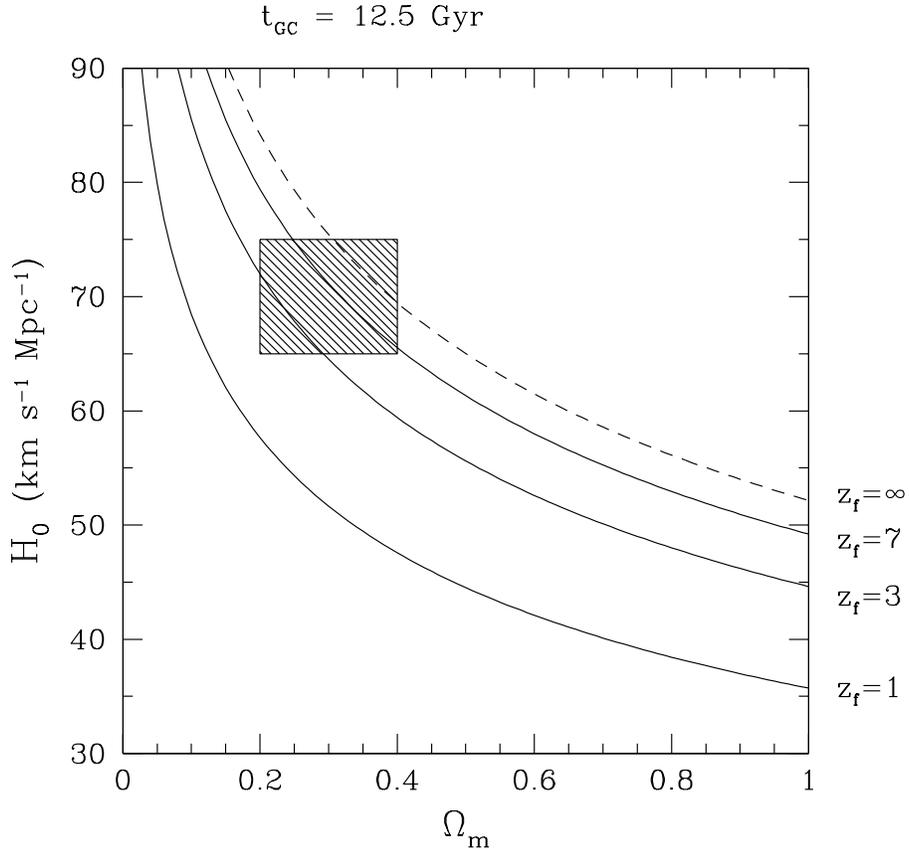}
\end{center}
\caption{The Hubble constant $H_0$ vs matter density $\Omega_m$,
required to give the age of expansion consistent with the redshift of
globular cluster formation $z_f$, assuming globular cluster ages are
fixed at $t_{GC} = 12.5$ Gyr and the Universe is flat (cf
Fig. \protect\ref{fig:age1}).  Dashed line shows the limiting case $t_H
= t_{GC}$.  Shaded box contains the 'best-fit' values.}
\label{fig:age2}
\end{figure}

Figure \ref{fig:age2} shows the constraints on $H_0$ and $\Omega_m$,
assuming that the absolute ages of globular clusters are fixed.  Again,
the current 'best-fit' values are consistent only with $z_f \gta 3$.

We have thus compared cosmic ages as estimated by researchers in three
different (somewhat unrelated) fields of astrophysics: the age of
expansion of the Universe, the age of globular clusters, and the epoch
of their formation.  We find that the three are consistent (within the
errors) only over a relatively small parameter space, and hence there
might be a problem with any one of the estimates.  This leads to several
possible conclusions:

(i) If the currently popular cosmological model ($\omegam=0.3,
\omegal=0.7, h=0.7$) is correct and the age estimates of globular
clusters are reliable, then one can put an extra constraint on the
models of globular cluster formation, such that $z_f \gta 3$.

(ii) If we trust the cosmology and wish to have a late globular cluster
formation, then a revision of the globular cluster ages (which are
somewhat model-dependent) is required.

(iii) If future research suggested that even old globular clusters
formed at $z_f \lta 3$ and the age estimates of globular clusters are
valid, then the above comparison indicates a problem for the `standard'
$\Lambda$-CDM model.  This is because the time-lag between the Big Bang
and the era of globular cluster formation would then be $> 2.1$ Gyr,
larger than the estimated errors.  The constraint is even more severe if
the ``dark energy'' is time-dependent quintessence rather than a
constant $\Lambda$.  Our overall conclusion is that there is a strong
cosmological (as well as astrophysical) motivation for firming up our
understanding of globular cluster formation.

\begin{table*}
\caption{\sc Scoreboard of globular cluster models}
\begin{center}
\begin{tabular}{lcccc}
\hline\hline
Observational properties           &  P\&D   & Hierarchy &  F\&R   & Merger \\
\hline
1. Narrow range of metallicity within \\
\quad individual clusters ($\delta\mbox{[Fe/H]} \lta 0.1$)
                                   & \check  & \check   & \check  & \check \\
2. Average metallicity $Z \sim 0.03\, Z_{\sun}$
                                   & X$^a$   & \check   & ?$^b$   & X$^c$ \\
3. Spherical geometry of \\
\quad globular cluster system
                                   & \check  & \check   & ?-X$^d$ & X$^e$ \\
4. Low efficiency $\epsilon_{GC}$
                                   & X$^f$   & \check   & \check  & ?$^g$ \\
5. Spread of relative ages of \\
\quad oldest clusters $\sim 2$ Gyr
                                   & X       & ?$^h$    & \check  & \check \\
6. Young massive clusters have \\
\quad a broad mass function
                                   & X$^i$   & \check   & ?$^j$   & ?$^j$ \\
7. Globulars have similar properties \\
\quad regardless of the size or morphology \\
\quad of the host galaxy
                                   & \check  & \check   & ?       & ? \\
\hline
Overall$^k$                        & X       & \check   & ?       & X \\
\hline
\smallskip
\end{tabular}
\begin{tabular}{ll}
  $a$ & in this model globular clusters form immediately after \\
      & recombination when the intergalactic gas has few metals; \\
      & self-enrichment is unlikely as globular clusters must \\
      & collapse quickly enough to remain gravitationally-bound \\
  $b$ & also likely to have lower metallicity; pre-enriched gas \\
      & would have cooled to a temperature lower than $10^4$ K \\
  $c$ & remaining gas in normal spiral galaxies is already enriched, \\
      & $Z \sim 0.1 - 1\, Z_{\sun}$, overproducing metallicity in new GCs \\
  $d$ & spherical geometry is assumed, but cloud fragmentation is \\
      & likely to proceed in highest density regions, i.e. in the disk \\
  $e$ & gas in mergers is contained within the orbital plane, \\
      & and new GCs must have a highly flattened distribution \\
  $f$ & this model assumes almost 100\% efficiency of gas conversion \\
  $g$ & observations suggest a possibly higher efficiency \\
  $h$ & age spread could be attributed to different time of virialization \\
      & of the progenitors of different mass and the statistical variance \\
      & of the progenitors of similar mass, creating a range of \\
      & redshifts of formation \\
  $i$ & this model reproduces only the median mass scale,
        $M \sim 10^5\, M_{\sun}$ \\
  $j$ & in these models cluster mass would depend on the virial temperature \\
      & of the halo and the metallicity of cooling gas \\
  $k$ & note that these marks apply only to the formation of globular \\
      & clusters in the Galaxy.
\medskip\\
\hline\hline
\end{tabular}
\end{center}
\end{table*}


\bigskip
Correspondence and requests for materials should be addressed to O.Y.G.
(e-mail: ognedin@ast.cam.ac.uk).

\end{document}